\documentclass[12pt]{article}
\usepackage{graphics}
\usepackage{cite,a4wide}
\newcommand{\beq}{\begin{equation}}
\newcommand{\eeq}{\end{equation}}
\newcommand{\bea}{\begin{eqnarray}}
\newcommand{\eea}{\end{eqnarray}}
\renewcommand{\d}{\delta}

\renewcommand{\b}{\beta}

\renewcommand{\ni}{\noindent}

\newcommand{\m}{\mu}

\newcommand{\s}{\sigma}

\renewcommand{\th}{\theta}

\newcommand{\oh}{\frac{1}{2}}

\newcommand{\dg}{\dagger}
\newcommand{\non}{\nonumber}

\newcommand{\rf}[1]{(\ref{#1})}
\newcommand{\ra}{\rightarrow}

\begin{document}

\hfill October 1998

\begin{center}

\vspace{32pt}

  { \bf \large Center Projection With and Without Gauge Fixing }

\end{center}

\vspace{18pt}

\begin{center}
{\sl M. Faber${}^a$, J. Greensite${}^{bc}$,
and {\v S}. Olejn\'{\i}k${}^d$}

\end{center}

\vspace{18pt}

\begin{tabbing}

{}~~~~~~~~~~~~~~~~~~~~\= blah  \kill
\> ${}^a$ Inst. f\"ur Kernphysik, Technische Universit\"at Wien, \\
\> ~~A-1040 Vienna, Austria.  E-mail: {\tt faber@kph.tuwien.ac.at} \\
\\
\> ${}^b$ Physics and Astronomy Dept. San Francisco State Univ., \\
\> ~~San Francisco, CA 94117 USA  E-mail: {\tt greensit@stars.sfsu.edu} \\
\\
\> ${}^c$ Theory Group, Lawrence Berkeley National Laboratory, \\
\> ~~Berkeley, CA 94720 USA  E-mail: {\tt greensit@lbl.gov} \\
\\
\> ${}^d$ Institute of Physics, Slovak Academy of Sciences, \\
\> ~~SK-842 28 Bratislava, Slovakia.  E-mail: {\tt fyziolej@savba.sk}

\end{tabbing}

\vspace{18pt}

\begin{center}

{\bf Abstract}

\end{center}

\bigskip

   We consider projections of SU(2) lattice link variables
onto $Z_2$ center and U(1) subgroups, with and without gauge-fixing.  
It is shown that in the absence of gauge-fixing, and up to an additive
constant, the static quark
potential extracted from projected variables agrees \emph{exactly} with 
the static quark potential taken from the full link variables; this is an
extension of recent arguments by Ambj{\o}rn and Greensite, and by
Ogilvie.  Abelian and center dominance is essentially trivial in this
case, and seems of no physical relevance.  The situation changes
drastically upon gauge fixing.  In the case of center projection, there
are a series of tests one can carry out, to check if vortices identified in the
projected configurations are physical objects.  All these criteria are
satisfied in maximal center gauge, and we show here that they all fail in the
absence of gauge fixing.  The non-triviality of center projection is due
entirely to the maximal center gauge-fixing, which pumps information about the
location of extended physical objects into local $Z_2$ observables.
\vfill

\newpage

\section{Introduction}

   There is currently a debate in the lattice gauge
theory community regarding which type of gauge field configuration is
responsible for the confining force.  Of course, confinement mechanisms 
involving monopoles, center vortices, instantons, and various other types of 
topological objects 
have been long discussed in the literature, over a period of 
decades.  In recent years, however, some of these proposals are being 
subjected to numerical tests.  In this connection, it is useful to ask if 
all of these tests really give us new information, or if, instead, certain 
results turn out as they do for some very trivial reason.

   In this article we will be largely concerned with the center vortex
theory, and with the (somewhat vague) concept of ``center dominance,'' 
and most especially with the ability of center projection to identify 
physical objects in the vacuum.  Much of the discussion, however, applies to 
abelian dominance and abelian projection as well.   

   We will begin by showing, in section 2, that in the absence of
gauge-fixing, and apart from an additive constant, the potential extracted 
from center-projected and/or abelian-projected 
lattices agrees \emph{exactly} with the potential derived from the unprojected
lattice; i.e.\ not only at large distances, but also in the Coulomb regime.  
If this is what is meant by center or abelian dominance,
then it is essentially a triviality in the absence of gauge-fixing, 
having no obvious relevance to the physics of confinement. This result
is an extension of remarks by Ambj{\o}rn and one of the authors
\cite{GG3}, and of recent work by Ogilvie \cite{mog}.

   The situation is much different when gauge-fixing is imposed. 
Center projection in maximal center gauge has been used to identify the 
location of center vortices, and it is asserted that these are physical 
objects.  In ref.\ \cite{PRD98}   
we reported the results of a series of numerical tests demonstrating
the physical nature of vortices identified in center projection; these
include such things as the effect of vortices on large, unprojected 
Wilson loops, asymptotic scaling of the vortex density, and other properties
discussed below. It is a compelling illustration of the importance of maximal 
center gauge-fixing to simply repeat the numerical tests in the absence of 
gauge-fixing.  What we find, in section 3, is that every one of these tests
of ``physicality'' 
fails, when no gauge-fixing is employed.  This failure is not at all 
surprising.  Vortices are located using local operators (the 
center-projected plaquettes), and in the absence of a global gauge-fixing 
these operators can hardly be expected to contain information about
infrared physics.  But the failures of the no gauge-fixing case serve
to highlight the remarkable, and highly non-trivial, fact that each test is
satisfied when maximal center gauge is imposed.

   Finally, in section 4, we note that the ``center dominances'' which
are obtained with and without gauge-fixing are not really the same.  
Without gauge-fixing,
projected and unprojected potentials are identical, starting out Coulombic at
short distances and going linear at large distances.  Imposing 
maximal center gauge, the center-projected potential is nearly linear 
everywhere, from one lattice spacing onwards.  We explain why this 
``precocious linearity'' is to be expected, if projected plaquettes locate 
the genuine confining configurations (center vortices)
in the unprojected lattice.  Section 5 contains some concluding remarks.

\vspace{33pt}

\section{Center Dominance Without Gauge Fixing}

   Let $U(C)$ denote the product of link variables around
loop $C$ in SU(2) lattice gauge theory, and let $U(R,T)$ in particular
denote the link product around a rectangular $R\times T$ loop.
Suppose, instead of computing the potential in the usual way, i.e.
\beq
       V(R) = \lim_{T\ra \infty} - \log\left[ {<\mbox{Tr}[U(R,T+1)]> \over 
<\mbox{Tr}[U(R,T)]> }\right]
\eeq
we calculate the potential from only the sign of the Wilson loop
\beq
V_S(R) = \lim_{T\ra \infty} - \log\left[ {<\mbox{signTr}[U(R,T+1)]> \over 
<\mbox{signTr}[U(R,T)]> } \right] 
\eeq
This was done done numerically in ref.\ \cite{TK1} (although of course 
without taking $T$ to $\infty$), with the surprising result that the 
sign-projection potential $V_S(R)$ and the full potential $V(R)$ agree, 
except for the very smallest loops.  However, this agreement can actually 
be explained in a simple way, as shown in ref.\ \cite{GG3} 
(see also \cite{PS,Gop}).

   We first need the result that for large $T$,
\beq
      W_{1/2}[R,T] \gg W_{3/2}[R,T] \gg W_{5/2}[R,T] \gg ...
\label{ineq}
\eeq
where
\beq
      W_j[C] = {1\over 2j+1} <\chi_j[U(C)]>
\eeq
and $\chi_j[g]$ is the SU(2) group character in representation $j$.
The above inequality \rf{ineq} can be seen as follows.  In all cases, we
consider very large $T$, and $j=$half-integer.  Begin with $R$ in the 
Coulombic regime.
The leading contribution, just coming from 1-gluon exchange,
is
\beq
       W_j[R,T] \approx \exp\Bigl[-T\Bigl(-{g^2_{eff}(R) C_j \over 4\pi R} 
                + 2b g_{eff}^2(a) C_j\Bigr)\Bigr] 
\eeq
where $C_j=j(j+1)$ is the quadratic Casimir and $b$ is a
constant of $O(1)$.  The first term is the Coulomb contribution, the
second term is the self-energy, and we have neglected terms subleading
in $T$.  Because $C_j$ increases with $j$, and because the self-energy 
exceeds the Coulomb term, the inequality \rf{ineq}
at large $T$ follows.  As $R$ is increased, and the loop probes
forces in the Casimir-scaling regime, the leading loop behavior becomes
\beq
       W_j[R,T] \approx \exp\Bigl[-T\Bigl(\s_j R 
                + 2b g^2_{eff}(a)C_j\Bigr)\Bigr] 
\eeq
where the string tension $\s_j$ increases with $j$.  Here again, since
$\s_j$ and $C_j$ increase with $j$, the inequality \rf{ineq} is satisfied.
Finally, in the
asymptotic regime, the color charges of the half-integer representations
are screened via binding to gluons down to $j=\oh$, 
and we have 
\beq
       W_j[R,T] \approx \exp\Bigl[-T\Bigl(\s_{1/2} R 
                + d_j)\Bigr] 
\eeq
The term $d_j$ contains two contributions.  The first, for $j>\oh$, is the
bound state energy of gluons required to
screen the heavy quark color charge to $j=1/2$.  The higher $j$ is, the more 
gluons are required to screen the charge, and the larger the energy of the 
``gluelump.''  The constant $d_j$ also includes the heavy-quark perturbative 
self-energy contribution,
proportional to $g^2 C_j$.  Both contributions cause $d_j$ to
increase with $j$, and again \rf{ineq} is obtained.  The conclusion is that, 
for any $R$, and any two half-integer representations $j_1>j_2$, 
\beq
       \lim_{T\ra \infty} {W_{j_1}[R,T] \over W_{j_2}[R,T] } = 0
\label{lim}
\eeq  
We now make the character expansion
\beq
      \mbox{signTr}[g] = \sum_{j=\oh,{3\over 2},{5\over 2},...} a_j \chi_j[g]
\eeq
with
\bea
       a_{1/2} &=&  \int dg ~ \mbox{signTr}[g] \chi_{1/2}[g] 
\non \\
               &=& {8\over 3 \pi}
\eea
Then, taking into account \rf{lim}, this means that for large $T$
\beq
       <\mbox{signTr}[U(R,T)]> \approx {8 \over 3 \pi} <\mbox{Tr}[U(R,T)]>
\eeq
and the equality of the full potential $V(R)$ and projected potential
$V_S(R)$ follows immediately.  From its derivation, which simply 
follows from the character expansion and eq.\ \rf{ineq}, it is not clear to 
us that the equality of $V_S(R)$ and $V(R)$ bears directly on the 
confinement issue.  Note that this equality holds even in the Coulombic
regime, before confinement physics comes into play.

   A closely related observation has been made by Ogilvie 
\cite{mog}, this time 
concerning projections of link variables, rather than loop variables.  Consider
a projection $U_\m(x) \ra H_\m(x)$, of SU($N$) link variables onto some
subgroup $H$ of SU($N$). Then it is shown in ref.\ \cite{mog} that, in the
absence of gauge-fixing, the asymptotic
string tension extracted from the projected link variables in Tr$H(C)$
agrees with the asymptotic string tension derived from the full link
variables.

  In fact, for both abelian and center projection, the statement concerning 
potentials from full and projected link variables can be made very much 
stronger than the result stated in \cite{mog}.  We begin with
center projection.  The Wilson loop on a 
center-projected lattice is defined to be
\bea
         W_P(C) &\equiv& <\prod_{l\in C} \mbox{signTr}[U_l]>
\non \\
          &=& {1\over Z} \int DU \prod_{l\in C} 
    \sum_{j_l=\oh,{3\over 2},{5\over 2},...} a_{j_l} \chi_{j_l}[U_l] e^{-S}
\eea
Let $(x_l,y_l)$ denote the (path-ordered) endpoints of link $l\in C$, with 
the convention \newline
$U_l=U_\m(x_l)$ if $y_l = x_l+\hat{\m}$, and
$U_l=U^{\dg}_\m(y_l)$ if $x_l = y_l+\hat{\m}$.  
Applying the familiar trick
of inserting an integration over gauge transformations $1=\int Dg$ 
followed by a change of variables $U\ra gUg^\dg$, 
\bea
     W_P(C)  &=& {1\over Z} \int DU Dg \prod_{l\in C} 
   \left(  \sum_{j_l=\oh,{3\over 2},{5\over 2},...} a_{j_l} 
          \chi_{j_l}[g(x_l)U_lg^\dg(y_l)] \right) e^{-S}
\non \\
     &=& {1\over Z} \int DU 
          \sum_{j=\oh,{3\over 2},{5\over 2},...} a_j^{P(C)}
          {1\over d_j^{P(C)-1}} 
          \chi_j[U(C)] e^{-S}
\non \\
     &=& \sum_{j=\oh,{3\over 2},{5\over 2},...} a_j^{P(C)}
          {1\over d_j^{P(C)-2}} W_j(C) 
\eea
where $P(C)$ is the loop perimeter, and we have used the identity
\beq
     \int dg \chi_j[Ug^\dg] \chi_k[gU'] = {1\over d_j} \chi_j[UU'] \d_{jk}
\eeq
with $d_j=2j+1$.  Finally, making use of \rf{lim}
we have 
\beq
       W_P(R,T) \stackrel{T\ra \infty}{\Longrightarrow} 
       4 \left({4\over 3\pi}\right)^{P(C)} W_{1/2}(R,T) 
       ~~~~ \mbox{\bf (center projection)}
\label{p1}
\eeq
and the equality of center-projected and full potentials up to an
additive constant 
\beq
          V_P(R) + 2\ln\Bigl({4\over 3\pi}\Bigr) = V(R)
\label{vpv}
\eeq
again follows immediately.  

   Abelian links $A$ are obtained from full link variables $U$
\beq
       U = \left( \begin{array}{cc}
         \cos\phi ~ e^{i\th}   & \sin\phi ~ e^{i\chi} \cr
        -\sin\phi ~ e^{-i\chi} & \cos\phi ~ e^{-i\th}   \cr
            \end{array} \right)
\eeq
by setting
\beq
      A(U) = \left( \begin{array}{cc}
           e^{i\th}   &  0          \cr
             0        &  e^{-i\th}   \cr
            \end{array} \right)
\eeq
and the abelian Wilson loop is defined as
\beq
      W_P[C] = \oh <\mbox{Tr}[\prod_{l\in  C} A(U_l)]>
\eeq
Abelian links, unlike center projected links, are not class functions,
and cannot be expanded in SU(2) group characters.  Instead we make use
of the fact that, in the absence of gauge-fixing, only the gauge-invariant 
component $O_{inv}$ of an operator $O$ will contribute to the expectation
value of $O$, where $O_{inv}$ is given by
\beq
     O_{inv}[\{U_l\}] = \int Dg ~ O[\{g(x_l)U_lg^{\dg}(y_l)\}]
\eeq
This fact is readily verified by again making the insertion $1=\int Dg$ and
change of variables $U\ra gUg^\dg$ in the functional integral, which 
gives, in the present case,
\beq
     W_P[C] = {1\over Z} \int DU \left\{ \int Dg ~ 
   \oh \mbox{Tr}[\prod_{l\in  C} 
    A(g(x_l)U_lg^\dg(y_l))] \right\} e^{-S}
\eeq
The quantity in braces is gauge invariant, and can now be expanded
in SU(2) group characters
\beq
\int Dg ~ \oh \mbox{Tr}[\prod_{l\in  C} A(g(x_l)U_lg^\dg(y_l))]
             = \sum_j a_j \chi_j[\prod_{l\in  C} U_l]
\eeq
where
\bea
      a_j &=& \int \prod_{l\in C}dU_{l}€ 
\int Dg ~ \oh \mbox{Tr}[\prod_{l\in  C} 
              A(g(x_l)U_lg^\dg(y_l))] \chi_j[\prod_{l\in  C} U_l]
\non \\
          &=& \int \prod_{l\in C}dU_{l}€  ~ \oh \mbox{Tr}[\prod_{l\in  C} 
              A(U_l)] \chi_j[\prod_{l\in  C} U_l]
\eea
and in particular          
\bea
     a_{1/2} &=& \left\{\prod_{l\in  C} {1\over 2\pi^2} \int_0^{\pi/2} d\phi_l
     \cos\phi_l \sin\phi_l 
\int_{-\pi}^{\pi} d\th_l \int_{-\pi}^\pi d\chi_l \right\}
\oh \mbox{Tr} \prod_{l'\in  C} \left( \begin{array}{cc}
           e^{i\th_{l'}}   &  0          \cr
             0        &  e^{-i\th_{l'}}   \cr
            \end{array} \right)
\non \\
    & & \times \mbox{Tr} \prod_{l''\in  C}\left( \begin{array}{cc}
         \cos\phi_{l''} e^{i\th_{l''}}   & \sin\phi_{l''} e^{i\chi_{l''}} \cr
        -\sin\phi_{l''} e^{-i\chi_{l''}} & \cos\phi_{l''} e^{-i\th_{l''}}   \cr
            \end{array} \right)
\non \\
             &=& \Bigl( {2\over 3}\Bigr)^{P(C)}
\eea
Thus we can again express
\beq
     W_P[C] =  \sum_{j=\oh,{3\over 2},{5\over 2},...} 
                    a_j <\chi_j[\prod_{l\in  C} U_l]>
\eeq
For the same reasons as before, the higher representations can be
neglected as $T\ra \infty$, so that
\beq
       W_P(R,T) \stackrel{T\ra \infty}{\Longrightarrow} 
       2\left({2\over 3}\right)^{P(C)} W_{1/2}(R,T) 
       ~~~~ \mbox{\bf (abelian projection)}
\label{p2}
\eeq
from which the exact equality (up to an additive constant) of the 
abelian-projected and full potentials
follows.  It is interesting that, although the expressions 
\rf{p1} and \rf{p2} relating
$W_P(R,T)$ to $W_{1/2}(R,T)$ only hold exactly as $T\ra \infty$, 
we have found in numerical simulations that these are in fact very good 
approximations to even the smallest projected $1\times 1$ and
$1\times 2$ loops at, e.g., $\b=2.3$.
     
   The initial interest in abelian projection (and, more recently,
center projection) was sparked by the discovery that the string tension 
of abelian projected lattices scales, and agrees (at least approximately) 
with the full string tension \cite{SY}.
The triviality of this result, in the absence of gauge fixing, suggests
that no strong conclusions can be drawn from abelian and/or center dominance
alone.  However, results obtained from lattice projection do not
end with abelian or center dominance.
In particular, in the case of center projection in maximal
center gauge, we claim to be able to identify the location of
confining center vortices in the unprojected lattice and to show, via
a series of tests, that these are physical objects rather than artifacts of
the projection.   The existence of center dominance \rf{vpv} in the
absence of gauge-fixing then raises a natural question:  Are the tests
for the physical nature of vortices \emph{also} somehow trivial?  Suppose one
tries to identify center vortices without any gauge fixing, and repeats
the same series of tests.  What happens?     

\section{Can One Find Vortices Without Gauge-Fixing?} 
             
   The ``direct'' version of maximal center gauge, used in most of our
work, is defined as the gauge which maximizes
\beq
       \sum_{x,\m} \Bigl| \mbox{Tr}[U_\m(x)] \Bigr|^2
\label{max}
\eeq
There is also an ``indirect'' version, which begins from
maximal abelian gauge and then uses the remnant $U(1)$ symmetry to maximize
an expression like \rf{max}, with the full link variables replaced by the 
abelian projected links.  Center projection (in SU(2) gauge theory) is a 
mapping of the SU(2) lattice link variables $U_\m$ to $Z_2$ link variables
$Z_\m$ via
\beq
       Z_\m(x) \equiv \mbox{signTr}[U_\m(x)]
\eeq
The excitations of a $Z_2$ lattice are $Z_2$ vortices, which are line-like
in D=3 dimensions, and surface-like in D=4 dimensions.  We refer to the
vortices on the center-projected lattice as projected vortices, or just
``P-vortices.''  A plaquette on the original lattice is said to be
``pierced'' by a P-vortex if the corresponding plaquette on the projected
lattice has the value $-1$.  We define vortex-limited Wilson loops $W_n(C)$
to be Wilson loops evaluated on the original, unprojected lattice, with
the following ``cut'' in the Monte Carlo data:  $W_n(C)$ is only evaluated
for those loops $C$ in which exactly $n$ plaquettes in the minimal area
are pierced by P-vortices.  In the same way, $W_{even}(C)$ and $W_{odd}(C)$
refer to loops pierced by even and odd numbers of P-vortices, respectively,
and one can also define Creutz ratios $\chi_n(I,J)$, $\chi_{even}(I,J)$
extracted from the $W_n$ and $W_{even}$.

   We now list our reasons, which are the outcome of a series of
tests, for believing that ``thin'' P-vortices in the center-projected lattice
locate ``thick'' center vortices in the unprojected lattice, and that
these thick vortices are physical objects:

\begin{itemize}

\item{ {\bf P-Vortices locate center vortices.~~} Vortex excitations
in the center-projected configurations, in direct maximal center gauge,
locate center vortices in the full, unprojected lattice.  The
evidence for this comes from the fact that 
\beq
   W_n(C)/W_0(C) \ra (-1)^n ~~~ \mbox{and} ~~~ 
   W_{odd}(C)\ra -W_{even}(C) 
\eeq
as loop area increases.}

\item{ {\bf No vortices $\Rightarrow$ no confinement.~~}  When Wilson
loops in SU(2) gauge theory  are evaluated in subensembles of 
configurations with no vortices (or only an even number of vortices) 
piercing the loop, the string tension disappears; i.e.
\beq
       \chi_0(I,J) \ra 0 ~~~,~~~ \chi_{even}(I,J) \ra 0
\eeq
as loop area increases.} 

\item{ {\bf Vortex density scales.~~} The variation of P-vortex density
with coupling $\b$ goes exactly as expected for a physical quantity with 
dimensions of inverse area \cite{Lang,PRD98}.}  

\item{ {\bf P-vortices are strongly correlated with action density.~~}
Plaquettes which are pierced by P-vortices have a very much higher 
\emph{unprojected} plaquette action than the vacuum average.    
Monopole loops lie on P-vortices \cite{Zako}.}  

\end{itemize}

\ni Regarding this last point, it is also found that monopoles, 
identified in the maximal abelian gauge, lie along center vortices,
found in the indirect maximal center gauge, in a monopole-antimonopole 
chain.  The non-abelian field strength of monopole cubes, above the lattice 
average, is directed almost entirely along the associated center vortices. 
Monopoles appear to be rather undistinguished regions of vortices, and may 
simply be artifacts of the abelian projection, as explained in ref.\ 
\cite{Zako}. 

   Finally, there is

\begin{itemize}

\item{ {\bf Precocious Linearity.~~} There is no Coulomb potential on
the center-projected lattice at short distances.  The projected potential is
linear from the beginning, with a string tension in agreement with
the one extracted asymptotically on the unprojected lattice.}

\end{itemize}

   As far as the ``precocious linearity'' of the projected potential
is concerned, we know already, from the validity of eq. \rf{vpv} at
all distances, that this result is not obtained in the absence of
gauge fixing.  A discussion of this point will be deferred to the next
section.  We will now present data on the other criteria, taken from
Monte Carlo simulations with no gauge fixing,
and compare it to our previous results, in which
the maximal center gauge was imposed.

\begin{figure}
\centerline{\scalebox{.7}{\includegraphics{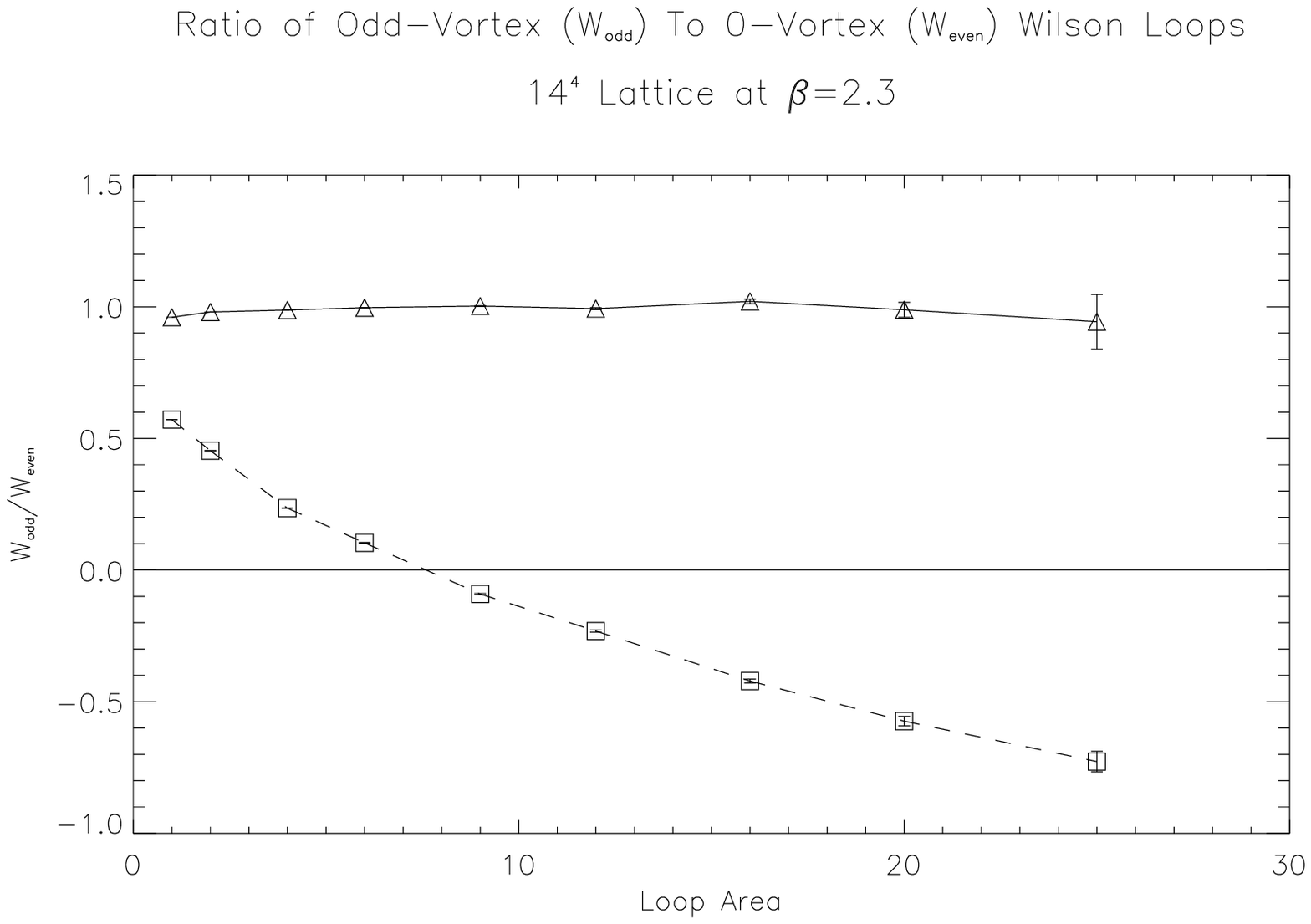}}}
\caption{Ratio of the Odd-Vortex to the Even-Vortex Wilson loops, 
$W_{odd}(C)/W_{even}(C)$, vs.\ loop area at $\b=2.3$, with and without
maximal center gauge-fixing.}
\label{mogvtex}
\end{figure}

   The first check is whether P-vortices still locate center vortices,
in the absence of maximal center gauge fixing.  The criterion (c.f.\
ref.\ \cite{PRD98}) is
that $W_n(C)/W_0(C) \ra (-1)^n$; however, since the density of P-vortices
is quite large in the absence of gauge fixing, it is hard to get
good statistics for $W_0(C)$ for the larger loops.  Instead, we
check the ratio $W_{odd}/W_{even}$.  If 
P-vortices locate center vortices, then we should also see
\beq
        {W_{odd}(C)\over W_{even}(C)} \ra -1
\eeq
as the loops get large.  The data at $\b=2.3$ on a $14^4$ lattice is
shown in Fig. \ref{mogvtex}.  The ratios obtained under maximal center gauge
are tending towards $-1$ as the loop area increases, which indicates that
P-vortices do indeed locate center vortices (we have shown $W_1/W_0$ and
$W_2/W_0$ data elsewhere \cite{PRD98}, this is also consistent with P-vortices
locating center vortices).  In obvious contrast, the ratio obtained
without gauge-fixing is consistent with $+1$, and has little variation
with loop size.  There is no reason, in this case, to suppose that
P-vortices locate center vortices.  So this is the first test to
fail in the no gauge-fix case.

\begin{figure}[h]
\centerline{\scalebox{.7}{\includegraphics{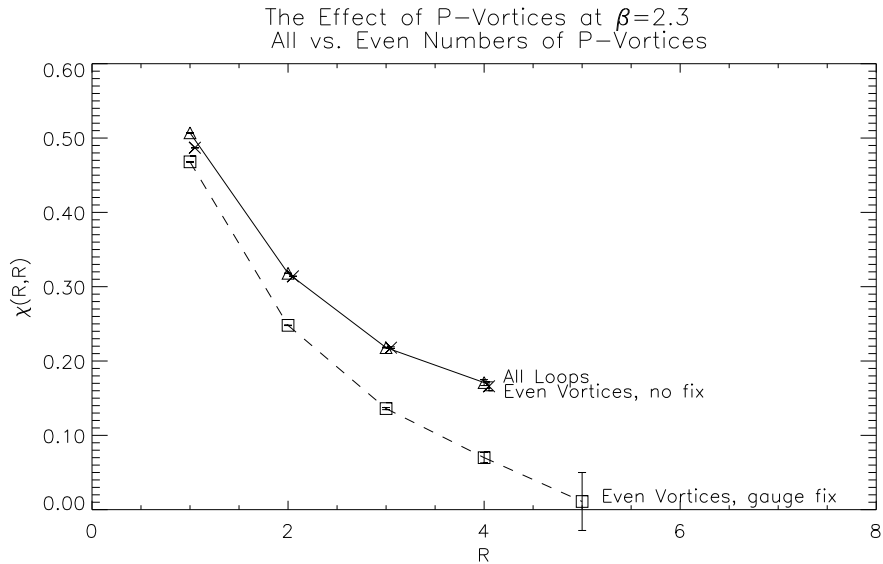}}}
\caption{Creutz ratios $\chi_{even}(R,R)$ extracted from loops pierced
by even numbers of P-vortices, defined with and without maximal center
gauge fixing, as compared to the usual Creutz ratios $\chi(R,R)$ at
$\b=2.3$. Data points
for $\chi_{even}(R,R)$ with no gauge-fixing (crosses) have been slightly 
displaced in $R$, to help distinguish them from the $\chi(R,R)$ data
points (triangles).}
\label{mogchieo}
\end{figure}

   The second test is to see if the no-vortex or even-vortex Wilson 
loops lose their string tensions, because of the zero- or even-vortex
restrictions.  Again,
because of statistics, we can only look at the even-vortex loops in
the no gauge-fix case. In Fig.\ \ref{mogchieo}
we see that in maximal center gauge the 
Creutz ratios $\chi_{even}(I,I)$ do indeed drop to zero as
loop area increases, while in the absence of gauge fixing there is
no discernable difference between  $\chi_{even}(I,I)$ and the usual
Creutz ratios $\chi(I,I)$.  

   Of course, the results shown in Figs. \ref{mogvtex} and \ref{mogchieo},
for the no gauge-fixing case, are closely related.  From Fig.\ 
\ref{mogvtex} we have that $W_{even}(C)\approx W_{odd}(C)$ in the no gauge-fix
case, which implies $W(C) \approx W_{even}(C)$.  The equality of Creutz
ratios $\chi_{even}(I,I)=\chi(I,I)$ follows, as seen in Fig.\
\ref{mogchieo}.

   The failure of these two tests, in the case of no gauge-fixing, 
could also have been anticipated analytically.  This will be shown in 
an appendix.

   For the test of asymptotic scaling, we first define 
$p$ to be the fraction, and $N_{vor}$ to be the total number, of center 
projected 
plaquettes with value $-1$.  $N_{vor}$ is also the total area of all 
P-vortices on the dual lattice, and we denote by $N_T$ the total number of 
all plaquettes on the lattice.  Then
\bea
           p &=& {N_{vor} \over N_T} = {N_{vor} a^2 \over N_T a^4} a^2
\non \\
             &=& {\mbox{Total Vortex Area} \over 
                  6 \times \mbox{Total Volume}} a^2
\non \\
             &=& {1 \over 6} \rho a^2
\eea
If $\rho$, which is the area of P-vortices per unit volume, 
is a fixed quantity in physical units, then according to 
asymptotic freedom we should find, in the scaling regime,
\beq
          p  = {1\over 6} {\rho \over \Lambda^2} 
                    \left( {6\pi^2 \over 11} \b \right)^{102/121} 
                    \exp\left[- {6\pi^2 \over 11} \b \right]
\label{p}
\eeq
where $a$ is the lattice spacing.  The fraction $p$ is related to
the 1-plaquette term in center projection
\beq
         W_{cp}(1,1) = (1-p) + p\times (-1) = 1-2p
\eeq

\begin{figure}[h]
\centerline{\scalebox{1.2}{\includegraphics{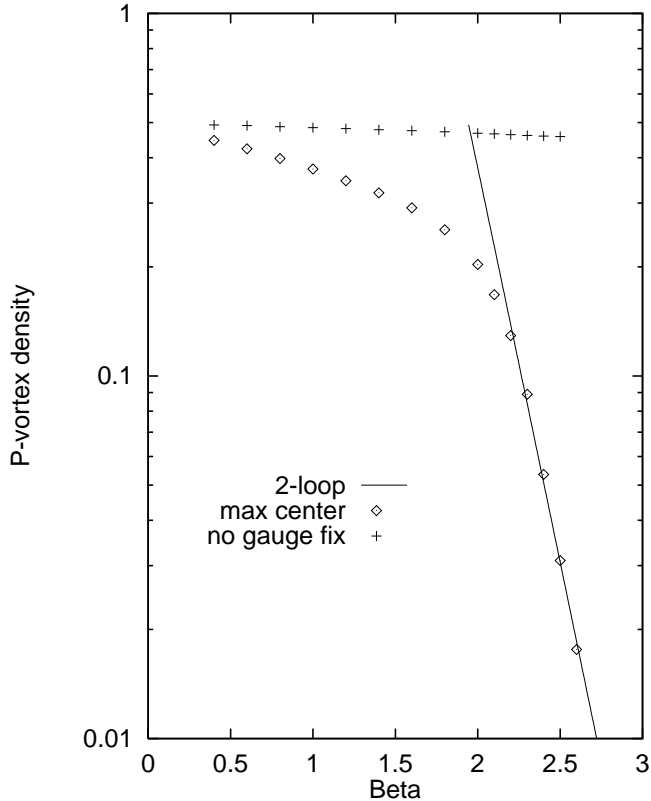}}}
\caption{Evidence for asymptotic scaling of the P-vortex density,
defined as the fraction $p$ of plaquettes pierced by P-vortices 
(one-sixth the average area occupied by P-vortices per unit lattice volume).
The solid line is the asymptotic freedom prediction of eq. \rf{p},
with constant $\protect \sqrt{\rho/(6\Lambda^2)} = 50$.  Data points
with and without maximal center gauge-fixing are shown.}
\label{pmog}
\end{figure}

   A plot of the P-vortex density $p$
versus coupling $\b$, as shown in Fig.\ \ref{pmog}.  The straight line
is the asymptotic freedom expression (last line of eq.\ \rf{p}), with 
the choice $\sqrt{\rho/(6\Lambda^2)} = 50$.  For the data obtained
in maximal center gauge, the scaling of P-vortex densities seems
really compelling (a result which was first reported for the
``indirect'' version of this gauge in ref.\ \cite{Lang}).
By contrast, the density of P-vortices obtained in
the absence of gauge-fixing shows no sign at all of asymptotic
scaling.

\begin{figure}[h]
\centerline{\scalebox{.9}{\includegraphics{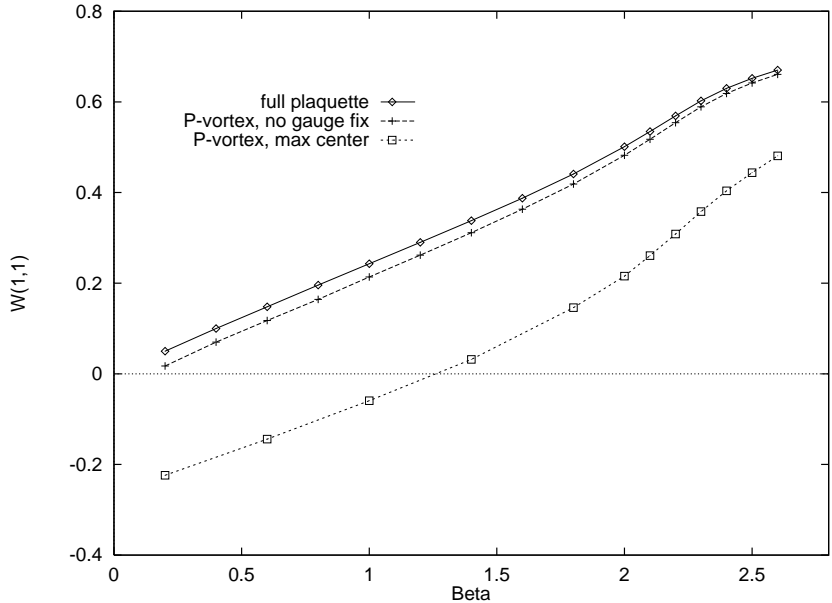}}}
\caption{One-plaquette loops $W_1(1,1)$ for plaquettes pierced
by P-vortices, evaluated both in maximal center gauge (squares) and with no 
gauge-fixing (crosses), as compared to the usual one-plaquette loop 
$W(1,1)$ (diamonds).}
\label{vort}
\end{figure}

   Finally we show in Fig.\ \ref{vort} a plot of the one-vortex plaquette
action $W_1(1,1)$ as a function of $\b$, both for maximal center gauge
and no gauge-fixing, compared to the usual plaquette action.  For no
gauge-fixing, the deviation of vortex plaquettes from average plaquettes
is very small.  In the maximal center gauge there is a very
substantial deviation.

   We conclude that the P-vortices identified in maximal center gauge
are true physical objects, which can be identified with center vortices.
In the absence of gauge fixing there is no indication, apart from
a very slight correlation with action density, that individual P-vortex
plaquettes identify the location of any physical object.

\section{Precocious Linearity}

   Only the gauge-invariant component $O_{inv}$ of an observable $O$,
obtained by averaging over gauge transformations,
contributes to the expectation value $<O>$, when the VEV is 
computed in the absence of gauge-fixing.  For abelian or 
center-projected Wilson loops,
the leading gauge-invariant contribution is the standard,
fundamental-representation Wilson loop, which in fact gives the entire 
contribution to the VEV in the $T\ra \infty$ limit. This is the 
quick explanation for why projected potentials agree with the full potential 
in the absence of gauge-fixing.  There is, however, no reason for such an 
agreement to persist when a global gauge-fixing is imposed, and in fact 
the exact agreement does not persist in general.

  In the first place, there is known to be 
a discrepancy between the string tensions 
of abelian projected loops obtained in maximal abelian gauge, and 
string tensions of unprojected loops. A careful study in lattice SU(2) gauge
theory at a particular coupling ($\b=2.5115$) shows that the string tension 
of projected loops in maximal abelian gauge is only $92\%$, rather than 
$100\%$ of the usual string tension \cite{Borny}.
There is also a difference between string tensions obtained in the ``direct''
\cite{PRD98} and ``indirect'' \cite{Us} versions of maximal center gauge,
with $\sqrt{\s}/\Lambda$ differing by $\approx 13\%$ in the two cases.
The projected string tensions in the direct version of maximal center gauge 
are in excellent agreement with the usual string tensions (see below), 
while the agreement in the indirect version is not so good.  
From these examples it is already apparent that when a global gauge-fixing 
is imposed, the equality of string tensions extracted from projected and 
unprojected loops is by no means guaranteed.  But a 
much more striking difference between the gauge-fixed and non-gauge-fixed cases
is found at short distances.

\begin{figure}[h]
\centerline{\scalebox{.7}{\includegraphics{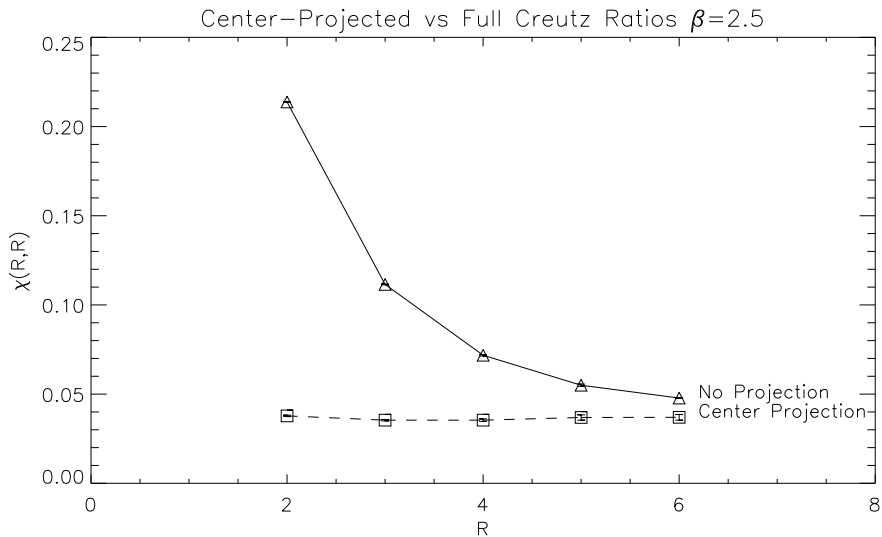}}}
\caption{Center-projected Creutz ratios $\chi_{proj}(R,R)$ in maximal
center gauge, as compared to the usual Creutz ratios $\chi(R,R)$ 
at $\b=2.5$.}
\label{mogchi}
\end{figure}

\begin{figure}[h]
\centerline{\scalebox{.6}[.9]{\includegraphics{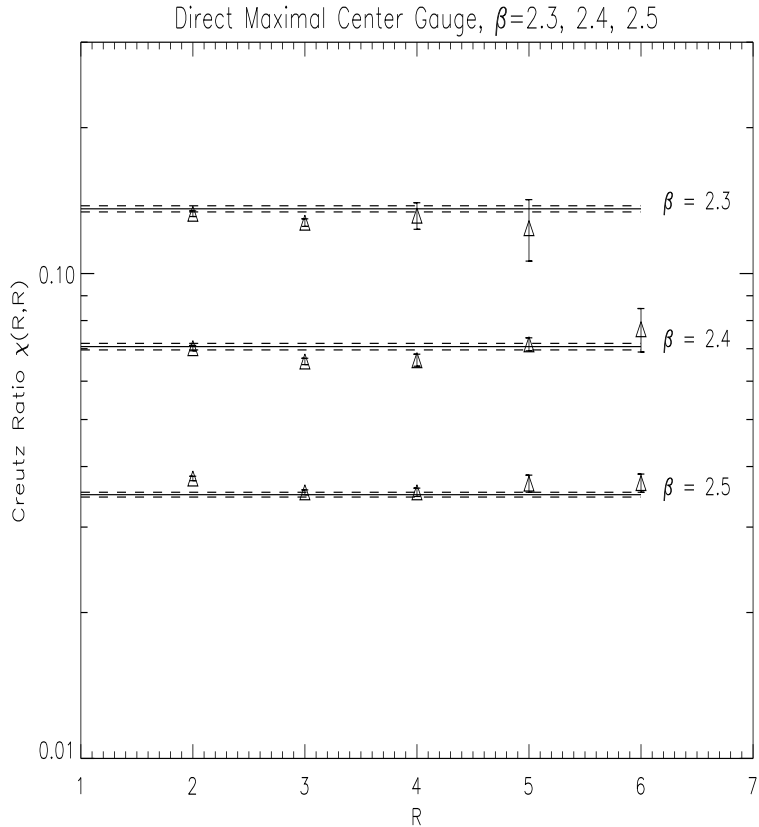}}}
\caption{Center-projection Creutz ratios $\chi(R,R)$ vs.\ $R$ at 
$\b=2.3,~2.4,~2.5$.  Triangles are our data points.  The solid line shows 
the value (at each $\b$) of the asymptotic string tension of the unprojected 
configurations, and the dashed lines the associated error bars, quoted in 
ref.\ \cite{Bali}.}
\label{ynot}
\end{figure}

   Since the abelian and center projected potentials agree \emph{exactly}
with the full potential (up to an additive constant) in the absence of
gauge-fixing, it follows that at short distances the projected potentials must
have a Coulombic form.  In contrast, Creutz ratios of the center-projected
lattice in maximal center gauge are basically
constant, starting with $\chi(2,2)$;
this means that the potential is linear starting at one lattice spacing.
This ``precocious'' linearity, for center projection in maximal center
gauge, is illustrated in Fig.\ \ref{mogchi}, where we compare center
projected Creutz ratios at $\b=2.5$, to the corresponding full Creutz ratios 
$\chi(R,R)$ quoted in ref.\ \cite{Gutbrod}.
A sampling of center projected Creutz ratios in maximal center gauge,
compared to the asymptotic string tension reported in ref.\ \cite{Bali},
is shown at various values of $\b=2.3,~2.4,~2.5$ in Fig. \ref{ynot}. 
This linearity of the center projected potential deep in the Coulombic regime
is clearly not just a small perturbation of the non-gauge-fixed result.

   The key point here is that, in maximal center gauge,
the distribution of P-vortices is precisely that of center vortices, which
are large, extended, physical objects.  Only this distribution, reflecting
long-range, confining physics, contributes to Wilson loops on the 
center-projected lattice.  Short-range fluctuations which, on the unprojected
lattice, are responsible for the Coulomb potential, are simply
removed by the projection.  In contrast, with no gauge fixing,
the distribution of P-vortices in some region of scale $R$ just
reflects the underlying fluctuations (including gaussian fluctuations) 
on the unprojected lattice at the scale $R$, and is only marginally affected,
at small scales, by the presence or absence of ``thick'' center vortices on
the unprojected lattice.  This is why the 
projected potential, with no gauge-fixing, only recovers the Coulomb 
potential at short distances.   

   A similar effect is found in a variant of abelian projection
in maximal abelian gauge, known as ``monopole dominance.''  Monopoles are 
first located on abelian-projected lattices by the DeGrand-Toussaint
criterion, and their contribution to the abelian Wilson loops is computed
using the lattice Coulomb propagator.  The potential extracted from such
``monopole'' Wilson loops then displays precocious linearity \cite{Borny}.
Here again, the distribution of monopoles is presumably governed by
long-range physics.  The magnetic fields assigned to those monopoles by
the ``monopole-dominance'' procedure do not include the high-frequency 
field fluctuations responsible for the Coulomb potential; there are
only large-scale fluctuations resulting in a linear potential.  
    
  The relation between monopole and vortex distributions has been
discussed in ref.\ \cite{Zako}, where it was found that monopoles, identified 
in abelian projection, lie along center vortices, identified 
in center projection, in monopole-antimonopole chains. The 
gauge-invariant plaquette action around monopole cubes has also been computed,
and it is found that almost all the excess plaquette action, above the vacuum 
average, lies on plaquettes of the cube pierced by P-vortices.  Moreover, 
this action distribution is almost identical to that of any cube pierced by a 
P-vortex line, with no monopole inside.  This indicates that the monopoles
identified in abelian projection are rather undistinguished regions of
center vortices, as discussed in ref.\ \cite{Zako}.  Nevertheless, the 
fact that center vortices, under abelian projection in maximal abelian
gauge, appear as monopole-antimonopole chains, no doubt underlies the 
approximate agreement of the monopole-dominated and center-projected 
potentials (both of which display precocious linearity).

\section{Conclusions}

   We have shown that in the absence of gauge-fixing, potentials 
$V_P(R)$ and $V(R)$ obtained, respectively, from projected and unprojected 
Wilson loops, agree exactly (up to an additive constant) at all $R$.  
On the other hand, such agreement does not hold automatically when global 
gauge-fixing is imposed, particularly at short distances.  
In view of this, the criterion that $V_P(R)$ and $V(R)$ have the same 
asymptotic behavior should be viewed only as a \emph{necessary} condition
that configurations identified on projected lattices are physical objects,
with the required confinement properties.
Abelian and center dominance in themselves are by no means a sufficient
indicator of the physical nature of monopoles and/or 
vortices identified on projected lattices.  

   There are, however, a number of other tests which can be used to establish 
the physical nature of topological objects identified on projected lattices.
In the case of center vortices, where the vortices are identified
by projection in maximal center gauge, it was found that:
(i) if a Wilson loop is linked to a region which, according to the
center-projection, contains a vortex, then that Wilson loop picks up
a relative minus sign ($W_n/W_0\ra (-1)^n$); (ii) the presence of vortices 
in the projected lattice is correlated to the existence of a string tension 
on the unprojected lattice ($\chi_0(R,R)\ra 0$);
(iii) P-vortex densities scale as required by asymptotic freedom, 
in a way which is appropriate to a density of surfaces; and 
(iv) there is a very large excess plaquette action at plaquettes pierced 
by P-vortices.   Properties (i), (ii), and (iv) involve 
correlations between the P-vortices
of the projected lattice with gauge-invariant quantities (Wilson loops,
Creutz ratios, and plaquette actions, respectively) on the unprojected 
lattice, while property (iii) is required if P-vortices on the 
center-projected lattice correlate with the location of physical, 
surface-like objects on the unprojected lattice.  

   We have shown in this paper that all of the criteria (i-iv) above, for
the physical nature of vortices located via center projection,
fail completely if no gauge-fixing is imposed.  This is a simple
consequence of the fact that vortices are identified using local
operators (projected plaquettes), and in the absence of a global 
gauge-fixing these operators have no information about physics on
scales much larger than one lattice spacing.  The non-triviality of 
center projection, and its ability to locate confining center vortex
configurations, can be attributed entirely to the effect 
of fixing to maximal center gauge.  This does not mean that the
center vortex mechanism is in any sense gauge-dependent, and in fact much
of the data shown above concerns the effect of vortices on gauge-invariant
observables.  It is important to distinguish between the procedure 
for locating vortices, and the center vortices themselves; it is only our 
method for \emph{finding} vortices which relies on a gauge choice. 

   Finally, the fact that the center-projected potential has the 
correct string tension remains an important property of maximal center
gauge.  It is true that   
this property is a triviality in the absence of gauge-fixing, as shown
in ref. \cite{mog} and also here.  However, as noted in the last section, 
the agreement of projected and standard string tensions is neither trivial 
nor inevitable when a global gauge-fixing is imposed and 
``precocious linearity'' is obtained; this agreement is needed in order 
to establish that fluctuations of center vortices alone lead to the correct 
value of the string tension.  It is the tests of physicality (i-iv) above,  
\emph{combined} with the property of center dominance in maximal
center gauge, that together make the case for center vortices as the quark
confinement mechanism.

\vspace{33pt}
 
\ni {\Large \bf Acknowledgements}

\bigskip

   J.G. is happy to acknowledge the hospitality of the Niels Bohr
Institute, where some of this work was carried out.

   This work was supported in part by Fonds zur F\"orderung der
Wissenschaftlichen Forschung P11387-PHY (M.F.), 
the U.S. Department of Energy under Grant No. DE-FG03-92ER40711
and Carlsbergfondet (J.G.), and the Slovak Grant Agency
for Science, Grant No. 2/4111/97 (\v{S}. O.).  
   
\appendix

\vspace{33pt}

\section{Appendix}

   In this appendix we show why 
$W_{even}(C) \approx W_{odd}(C) \approx W(C)$, in
the absence of gauge fixing, up to very large distance scales.
If we denote by
\beq
       Z(C) = \prod_{l\in C} \mbox{signTr}[U_l]
\eeq
then it is not hard to see that
\bea
     W_{even}(C) &=& { <\oh (1 + Z(C)) \oh \mbox{Tr}[U(C)] >
               \over <\oh (1 + Z(C)) > }
\non \\
     W_{odd}(C) &=& { <\oh (1 - Z(C)) \oh \mbox{Tr}[U(C)] >
               \over <\oh (1 - Z(C)) > }
\eea
Applying eq.\ \rf{p1}, and also using the fact that
\beq
      \chi_{1/2}[U(C)]^2 = 1 + \chi_1[U(C)]
\eeq
we have
\bea
      W_{even}(C) &\approx& W(C) + 2\left({4\over 3\pi}\right)^{P(C)}
\non \\
      W_{odd}(C) &\approx& W(C) - 2\left({4\over 3\pi}\right)^{P(C)}
\label{Weo}
\eea
Since $W(C)$ has an area-law falloff, it is clear that when the area
of the loop is much larger than the perimeter, the first term on the
rhs, $W(C)$, can be neglected in comparison to the second term.  That
would give $W_{odd}/W_{even} \ra -1$, which is in apparent 
contradiction to the Monte Carlo results shown in Fig.\ \ref{mogvtex}.  
The paradox is resolved by computing how large the loop $C$ has to be, 
for this limiting ratio to be obtained.  Consider for simplicity,
square Wilson loops, and let $L$ be the length of a side such that
\beq
      W(L,L) = \left({4\over 3\pi}\right)^{4L}
\eeq
For loops of somewhat smaller area, we can neglect the second, rightmost terms
in eq.\ \rf{Weo}, so that $W_{even}\approx W_{odd}$.\footnote{For example, 
in the case of $\b=2.3$, we find $W(5,5) \approx 4 \times 10^{-4}$, while 
for this loop the second term $2(4/3\pi)^{P(C)}$ in eq. \rf{Weo}
is approximately $7 \times 10^{-8}$.} From
\beq
      W(L,L) = \exp[-\s L^2 - 4bg^2(a) L]
\eeq
we have
\bea
        L &=& {4\over \s} \Bigl(\log{3\pi \over 4} - bg^2(a) \Bigr)
\non \\
          &\approx& {4\over \s} \log{3\pi \over 4}
\eea
where we have used the fact that $g^2(a) \ra 0$ in the continuum,
$\b \ra \infty$ limit.  Since $\s(\b) \ra 0$ as $\b \ra \infty$,
it is clear that $L\ra \infty$ in lattice units in the same limit. 
Convert now to physical units
\beq
      L_p = L a ~~, ~~~~~~ \s_p = {\s \over a^2}
\eeq
and we find
\beq
      L_p = {4 \s_p^{-1} \log{3\pi \over 4} \over a }
\eeq
Since $a(\b) \ra 0$ goes exponentially to zero as $\b \ra \infty$,
it follows that $L_p$ diverges to infinity in \emph{physical}
units.  The conclusion is that in the continuum limit,
\beq
        W_{even}(C) = W_{odd}(C) = W(C)
\eeq
for all loops $C$, and, as a consequence,
\beq
        \chi_{even}(I,J) = \chi(I,J)
\eeq
everywhere, in the absence of maximal center gauge-fixing.
This explains the results shown above 
in Figs.\ \ref{mogvtex} and \ref{mogchieo}.

\end{document}